\documentclass[aip,reprint,amsmath,amssymb,floatfix,longbibliography,footinbib]{revtex4-1}

\usepackage{graphicx}
\usepackage{dcolumn}           
\usepackage{bm}                
\usepackage{subfigure}
\usepackage[normalem]{ulem}    
\usepackage{color}             
\usepackage{hyperref}          
\usepackage{natbib}

\newcommand{\bjerrum}{\lambda_\textmd{B}}

\draft 

\begin{document}

\title{Size Scaling of Neutral Polymers and Charged Polymers in Nanochannels}

\author{ Yu-Lin Lee$^{1}$ and Pai-Yi Hsiao$^{1,2}$}
\email[Corresponding author, email:\ ]{pyhsiao@ess.nthu.edu.tw}
\address{\small
$^{1}$Department of Engineering and System Science, National Tsing Hua University, Hsinchu, Taiwan, R.~O.~C.\\
$^{2}$Institute of Nuclear Engineering and Science, National Tsing Hua University, Hsinchu, Taiwan, R.~O.~C.}

\date{\today}

\begin{abstract}
We expand the blob theory for freely-jointed chains and perform molecular dynamics simulations to study the behavior of polymers confined in cylindrical channels. 
From weak to strong confinement, five scaling regimes, de Gennes, extended de Gennes, transition, backfolding, and Odijk regimes, are distinguished for neutral polymers.
The size scalings in each regime are derived as a function of the channel width.
The scaling exponents $-1$ and $-1/3$ are obtained for the transition and backfolding regimes, respectively, which result from the reduction of the excluded volume of the segments by restriction of the segment's orientation in the narrowed channels.
For charged flexible chains, the de Gennes regime is split into Flory-de Gennes and electro-de Gennes regimes owing to strong Coulomb repulsion in electrostatic blobs. 
Nonetheless, the extended de Gennes and transition regimes are shrunken. 
The study of the fluctuations of the chain size shows consistent scaling demarcations for both the neutral and charged chain systems.
\end{abstract}

\maketitle

When a polymer is confined in a channel, the variational behavior of chain size can be separated into different regimes, depending on the confinement \cite{Reisner2005, reisner2012dna, dai2016polymer}.
In weak confinement, the de Gennes regime has been validly explained by the blob theory over the years \cite{daoud1977statistics, de1979scaling} 
while in extremely strong confinement, the deflection chain theory has been used to understand the Odijk regime \cite{odijk1983statistics, odijk1984similarity}. 
What we know less well is the behavior in strong confinement.
Several regimes have been claimed and theories are still under development or debate.
For example, a chain can partially fold back when it leaves the Odijk regime by reducing the confinement.
This is called the backfolding regime and Odijk has proposed a one-dimensional Flory-type theory to explain it
\cite{odijk2008scaling}. 
Dai et al.~\cite{dai2012conformation} later suggested an alternative theory using a two-state cooperativity model of deflection segments and S-loops to predict its behavior.   
Recent study~\cite{Muralidhar2016} using Monte Carlo chain growth showed strong evidence in support of the previous theory. 
There was a debate for the existence of a regime, called the extended de Gennes regime, which is next to the de Gennes regime and shares a similar size scaling behavior.
Now the existence of the regime is broadly accepted \cite{tree2013extension, dai2014extended, Werner2014}.
Between these two regimes, a ``transition'' regime exhibits with a scaling exponent of $-1$, varying with the confinement width \cite{Wang2011}. 
Currently, no theory can well explain it \cite{reisner2012dna, dai2016polymer}. 

In this letter, we propose a simple, generalized scaling theory to describe the behavior of chain size, particularly in the region of strong confinement.
The theory is further verified by performing molecular dynamics simulations for neutral chains as well as for charged chains (polyelectrolytes).
The presence of electrostatic interactions and ions significantly changes the scaling behavior.
The divisions of regime are thus modified for charged chain systems.

We consider a polymer as a freely-jointed chain comprising $N$ self-avoided segments. 
Each segment is of length $\ell$ and width $w$. 
In dilute solutions, the scaling theory subdivides the chain into connected thermal blobs \cite{de1979scaling, rubinsteinpolymer}. 
Each blob envelopes $g_{\rm T}$ consecutive segments and pervades a spherical space of size  $\xi_{\rm T}$. 
By definition, the excluded-volume interaction energy of a thermal blob is equal to the thermal energy $k_{\rm B}T$. 
Inside the blobs exhibits ideal chain behavior, 
described by $\xi_{\rm T} \sim \ell g_{\rm T}^{1/2}$ where $g_{\rm T}\sim (\ell/w)^2$. 
The whole chain size is hence $R_{0} \sim \xi_{\rm T} (N/g_{\rm T})^{\nu}$, which can be derived to be $R_{0} \sim \ell^{2-2\nu} w^{2\nu-1} N^{\nu}$ with $\nu$ being the Flory exponent.

The polymer is now confined in a cylindrical channel of diameter $D$.
We discuss the variation of the chain size by decreasing $D$.
The confinement stars to influence the chain size when $D$ is smaller than $D^*_1\sim R_{0}$.
The chain is envisioned to break into a series of blobs of size $\xi_{\rm c}\sim D$, aligned along the channel, with the segments in the blobs behaving as the Flory chains. 
It can be shown that  the number of segments in a blob is $g_{\rm c} \sim \ell^{2-2/\nu} w^{1/\nu-2}D^{1/\nu}$ and the chain has a size of $R_1 \sim (N/g_{\rm c}) \xi_{\rm c} \sim N\ell^{2/\nu-2}w^{2-1/\nu}D^{1-1/\nu}$.
This is the famous de Gennes result \cite{de1979scaling,daoud1977statistics,rubinsteinpolymer}, which predicts a $D^{-2/3}$ scaling if $\nu$ takes the value $3/5$.
As $D$ decreases, $g_{\rm c}$ decreases and can be eventually smaller than $g_{\rm T}$.
It thus sets a lower boundary $D^*_2 \sim ~\ell^2/w$ for the ``de Gennes regime''.

When $D<D^*_2$, the blobs become the thermal blobs and are compressed to be ellipsoidal.
Let $\xi_{\rm c\parallel}$ and $\xi_{\rm c\perp}$ be the longitudinal and the transverse size of a compressed thermal blob, respectively. 
The confinement imposes the condition $\xi_{\rm c\perp}\sim D$ and we set the following three scaling equations:
\begin{eqnarray}
\xi_{\rm c\parallel}\sim C_{\parallel} \ell g_{\rm c}^{1/2}\\
\xi_{\rm c\perp}\sim C_{\perp} \ell g_{\rm c}^{1/2}\\
k_{B}T\frac{v_{\rm ex}g_c^2}{\xi_{\rm c\parallel}\xi_{\rm c\perp}^2} \sim k_{B}T
\end{eqnarray}
The first two equations describe the random-walk behavior of the chain segments in a blob.
$C_{\parallel}$ and $C_{\perp}$ are functions of $D$, $\ell$ and $w$, which are introduced to describe the anisotropy of the walk in the confined space.
The third equation equates the excluded-volume energy in a blob to the thermal energy where
$v_{\rm ex}$ is the effective excluded volume of a segment.
For large $D$, $v_{\rm ex}$ takes the bulk value $\ell^2 w$.
When $D<\ell$, the rotation of a segment is restricted by the channel wall.
$v_{\rm ex}$ thus decreases and can be written as $\ell^2 w\langle \sin\phi\rangle$ where $\phi$ is the angle between two segments.
If $\phi$ is small,  $\langle\sin\phi\rangle\sim D/\ell$ and $v_{\rm ex}$ scales as $\ell w D$.
The chain size derived from the three equations gives $R \sim (N/g_{\rm c}) \xi_{\rm c\parallel} \sim N\ell^2D^{-1}C_{\parallel}C_{\perp}$, 
subject to the constraint $C_{\parallel}C_{\perp}^3\sim v_{\rm ex}D/\ell^4$.

In this compressed thermal blob region, researchers have found two scaling behaviors \cite{reisner2012dna, dai2016polymer}.
The ``extended de Gennes regime'' predicts the chain size as $R_2\sim N\ell^{4/3} w^{1/3} D^{-2/3}$, 
using $v_{\rm ex}\sim\ell^2 w$.
We found that the $C$'s functions have  $C_{\parallel} \sim D^0$ and $C_{\perp}\sim (wD/\ell^2)^{1/3}$.
Simulations further showed the existence of a $D^{-1}$ scaling behavior, called ``the transition regime''.
It happens when  $D<D^*_3\sim \ell$ and we understood it as a result of the crossover of the excluded volume from $\ell^2 w$ to $\ell w D$.
We obtained $C_{\parallel} \sim \ell^{3/2}w^{1/2}D^{-1/2}v_{\rm ex}^{-1/2}$,
$C_{\perp}\sim \ell^{-11/6}w^{-1/6}D^{1/2}v_{\rm ex}^{1/2}$, and $R_3\sim N \ell^{5/3}w^{1/3}D^{-1}$.
The expressions of $R_2$ and $R_3$ are in agreement with the simulations of Muralidhar et al.~\cite{muralidhar2014backfolding}.
In Fig.~4 of the work, the chain extension $\langle X \rangle$ over the contour length $L$ was seen to scale as $(D/\ell)^{x}(w/\ell)^{1/3}$ with the exponent $x$ changing approximately from $-2/3$ in the extended de Gennes regime to $-1$ in the transition regime.

If $D$ decreases further more and has passed the transition regime,  $v_{\rm ex}$ becomes $\ell w D$ and we have 
$C_{\parallel} \sim (\ell/w)^{1/2}$, $C_{\perp}\sim \ell^{-7/6}w^{1/2}D^{2/3}$ and $R_4\sim N \ell^{4/3}D^{-1/3}$.
It predicts a new $D^{-1/3}$-scaling regime, which corresponds to the ``backfolding regime'' proposed recently by Odijk \cite{odijk2008scaling, dai2012conformation, muralidhar2014backfolding}.
The boundary with the transition regime is estimated at $D=D^*_4\sim \sqrt{\ell w}$.

\begin{figure}[htbp]
\begin{center}
  \includegraphics[width=0.30\textwidth,angle=270]{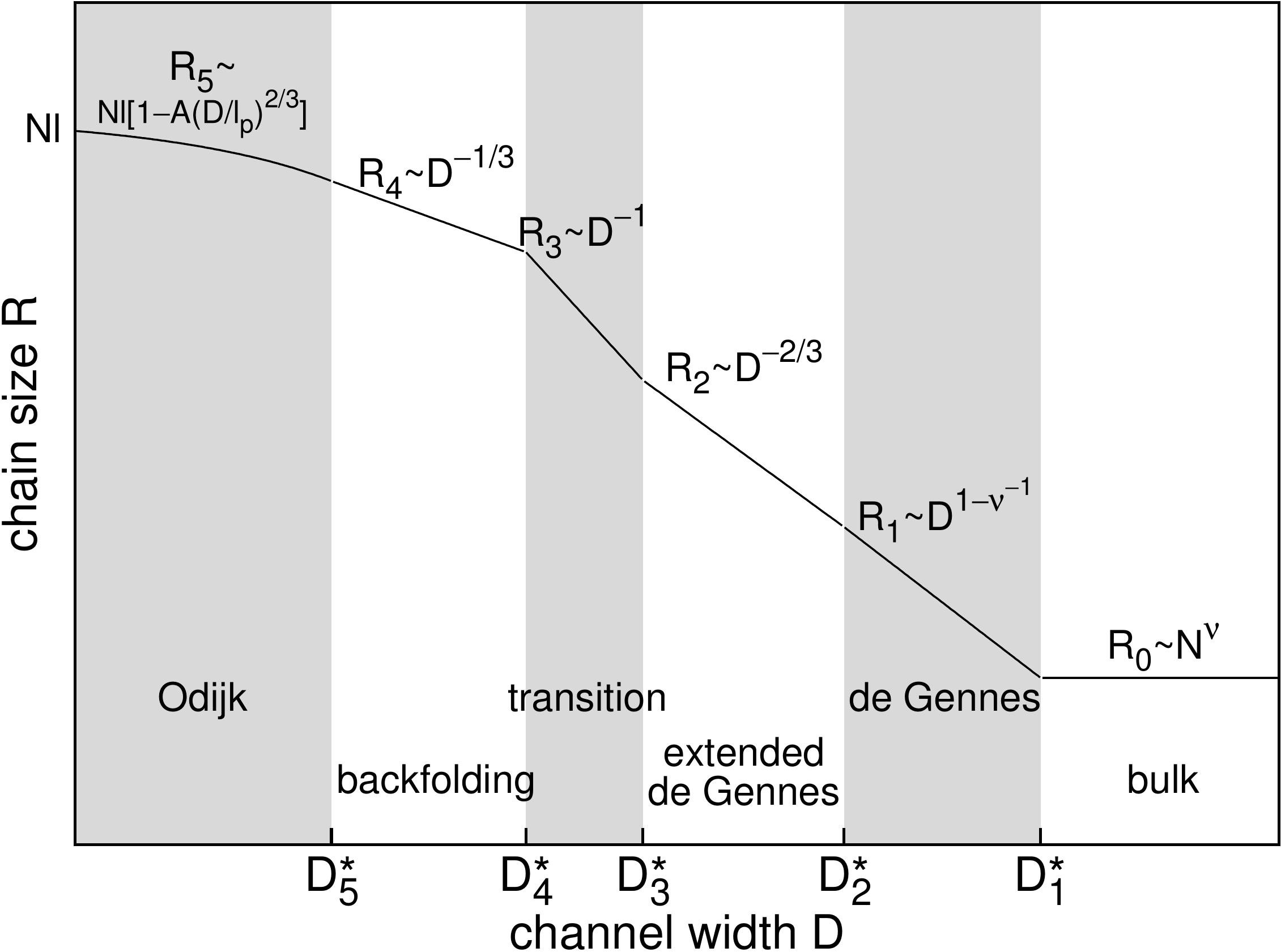}
  \caption{Scaling behavior of chain size $R$ as a function of channel width $D$ (plotted in log-log scales).
  The regime boundaries locate at $D^*_1\sim R_0$, $D^*_2\sim \ell^2/w$, $D^*_3\sim \ell$, $D^*_4\sim \sqrt{\ell w}$, and $D^*_5\sim \ell_{\rm p}$.
  }\label{fig:NCL_regimes}
 \end{center}
\end{figure}

For the extreme situation where the chain cannot fold back in a very narrowed channel, 
Odijk has formulated a deflection chain theory, using the wormlike chain model, to describe the chain size \cite{odijk1983statistics,odijk1984similarity}. 
The result is $R_5\sim N\ell [1- A (D/\ell_{\rm p})^{2/3}]$ where $\ell_{\rm p}$ is the persistence length and $A$ is a factor.
The regime boundary occurs at $D^*_5 \sim \ell_{\rm p}$.
The complete scaling behaviors for the chain size $R$ are summarized in Fig.~\ref{fig:NCL_regimes} as a function of the confinement width $D$. 

To verify the theory, Langevin dynamics simulations were performed to study single polymers confined in cylindrical channels, using bead-spring chain models. 
We first investigated neutral chain (NC) systems. 
Each monomer has mass $m$ and the excluded volume interaction was modeled by the Weeks-Chandler-Andersen (WCA) potential with the parameters $\varepsilon_{\rm m}=1.2k_{\rm B}T$ and $\sigma_{\rm m}=1\sigma$ where $k_{\rm B}T$,  $\sigma$, 
and $m$ are the energy, length, and mass units of simulation, respectively. 
To shorten the notations, we will describe the physical quantities by omitting these units in the following text. 
The monomers were bonded by a harmonic potential $u_{\rm b}(r)=\frac{1}{2}k_{\rm b}(r-b_0)^2$ with $k_{\rm b}=600.0$ and $b_0=1.0$. The chain stiffness was modeled by $u_{\rm a}(\theta)=k_{\rm a}(1-\cos(\theta-\pi))$ where  $\theta$ is the angle between two adjacent bonds. 
We simulated a semi-flexible chain with $k_{\rm a}=3.5$ and the number of monomers $N_{\rm m}=256$. 
We denote the chain `NC1'. 
The end-to-end distance of chain is $R_{\rm e}=42.6(8)$ in bulk solutions.
The persistence length is $\ell_{\rm p}=3.1(6)$, which is close to the theoretical estimate $b_0k_{\rm a}/(k_{\rm B}T)=3.5$. 
The model corresponds statistically to a Kuhn chain \cite{rubinsteinpolymer} with $\ell = 2 \ell_{\rm p} \simeq 6.2$, $w\simeq \sigma_{\rm m} =1$, and $N=N_{\rm m}b_0/\ell$.  
We confined the chain in a cylindrical channel of diameter $D_{\rm w}$. 
The monomer-wall interaction was modeled by a purely repulsive potential $u_{\rm w}(r)=\varepsilon_{\rm w}[\frac{2}{15}(\frac{\sigma_{\rm w}}{r})^9-(\frac{\sigma_{\rm w}}{r})^3]$ with $\varepsilon_{\rm w}=15$ and $\sigma_{\rm w}=(\sigma_{\rm m}+\sigma)/2$, cut off at the minimum point $r_{\rm c}=\sqrt[6]{2/5}\sigma_{\rm w}$. 
The above setting gives an effective channel width equal to $D=D_{\rm w}-1$. 
We varied the confinement from a large, quasi non-confined space, down to a strongly confined one. 
The results were further compared with two flexible chains ($k_{\rm a}=0$) of the same contour length:
the NC2 chain with $\sigma_{\rm m}=b_0=3.1$ and $N_{\rm m}=83$, and the NC3 chain with  $\sigma_{\rm m}=b_0=1.0$ and $N_{\rm m}=256$. 
These two cases have $\ell_{\rm p} \simeq w$.  
More information can be found in Supplemental Material.
    
Fig.~\ref{fig:nc_extension}(a) shows the variation of the chain extension $Z_{\rm ext}$ along the channel axis.  
\begin{figure}[htbp]
\begin{center}
\includegraphics[width=0.30\textwidth,angle=270]{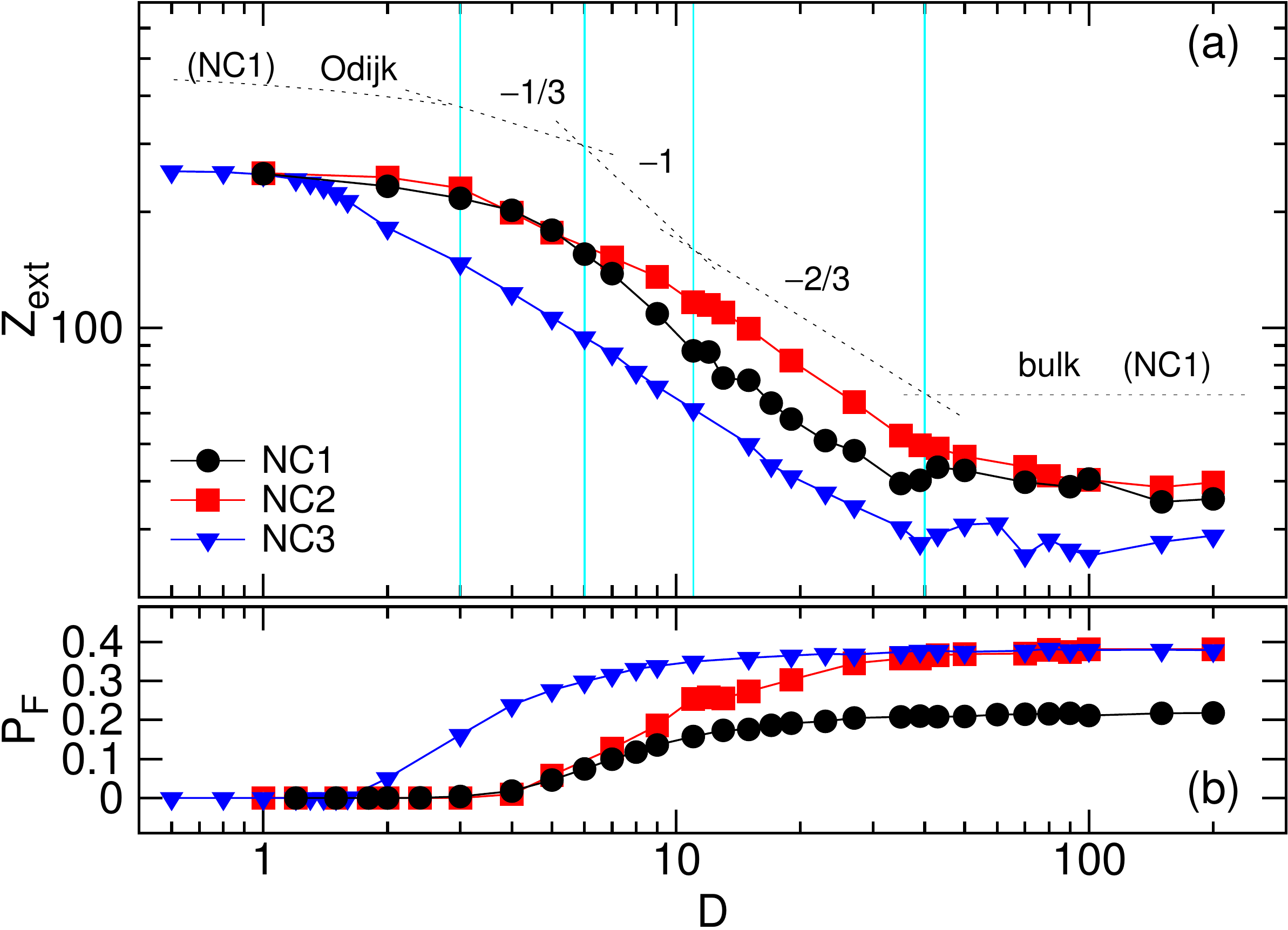}
\caption{(a) Chain extension $Z_{\rm ext}$ and (b) probability $P_{\rm F}$ of bond folding in the axial direction, as a function of $D$ for NC1, NC2, and NC3 chains. 
The scaling behaviors for NC1 chain have been indicated in dashed lines in Panel (a).
NC2 and NC3 chains have their own scalings and regimes (not indicated in the figure). }\label{fig:nc_extension}
 \end{center}
\end{figure}
We saw that $Z_{\rm ext}$ is about constant when $D$ is larger than $R_{0}$. 
As $D$ decreases to $D_{1}^*\simeq 40$, the extension starts to climb. 
The chain enters the de Gennes regime, and approximately scales as $D^{-2/3}$. 
Since the extended de Gennes regime shows a similar scaling behavior, we cannot simply determine the boundary with the de Gennes regime from the size variation.

For NC1, the extension curve shows the transition behavior ($D^{-1}$-scaling) in the region $6<D<11$. 
When $D<6$, the scaling exponent decreases and changes gradually to $-1/3$. 
The chain then enters the Odijk regime as $D<D_5^*\sim \ell_{\rm p}\sim 3$.
These boundaries agree with the predictions within prefactors which were ignored in the scaling derivations.
To show that the $D^{-1/3}$ scaling corresponds to the backfolding regime, 
we have calculated in Fig.~\ref{fig:nc_extension}(b) the probability $P_{\rm F}$ of bond folding in the axial direction.
$P_{F}$ is zero in the Odijk regime because the chain is unfolded.
It becomes non-zero when $D>3$, showing the occurrence of the chain backfold.

For NC2 and NC3 chains, we have $w\sim \ell$. 
It implies $D_3^*\sim D_4^*$, which consequently shrinks the transition regime. 
The chains now go from the de Gennes regime, directly through the backfolding regime, to the Odijk regime \footnote{To have $D_4^*>D_5^*$, $\ell$ must be smaller that $4w$. Therefore,
the theory also predicts a shrinkage of the backfolding regime when $\ell/w$ is large. 
It is because the high cost of the bending energy hinders the backward folding of the chain segments in a narrowed channel.}.
Compared with the NC1 chain, these regimes occur at different places and 
the probability $P_{\rm F}$ acquires a higher value because of the lack of the bending energy.
The effect of finite chain length has been verified in Fig.~S1.
The weak effect asserts that the scalings reported here have reached their asymptotic behavior.

\begin{figure}[htbp]
\begin{center}
  \includegraphics[width=0.30\textwidth,angle=270]{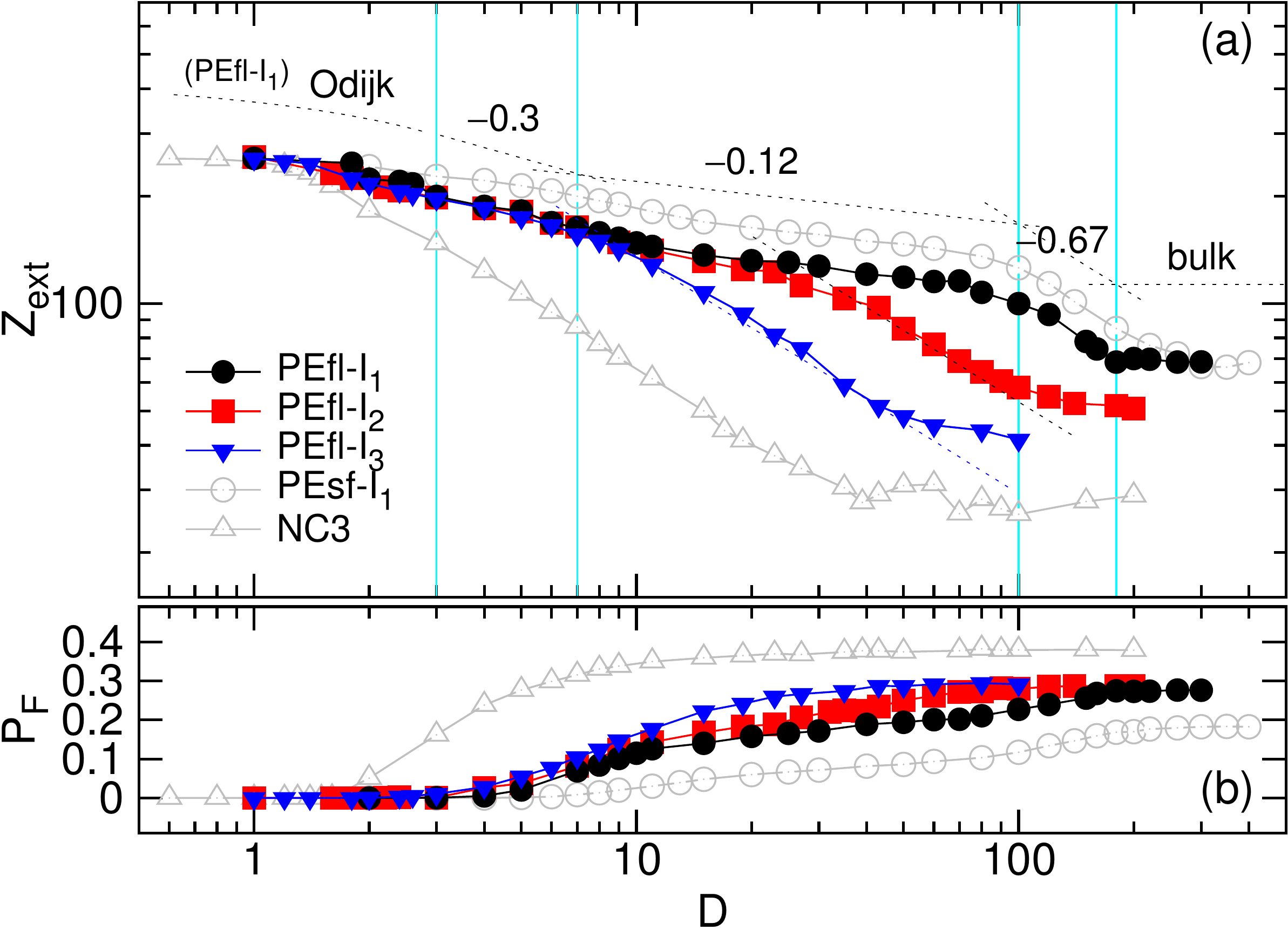}
  \caption{(a) Chain extension $Z_{\rm ext}$ and (b) probability $P_{\rm F}$ of bond folding in the axial direction, as a function of $D$ for PEfl-$I_1$, PEfl-$I_2$, PEfl-$I_3$, PEsf-$I_1$, and NC3 chains. 
  The scaling behaviors for PEfl-$I_1$ chain have been indicated in dashed lines in Panel (a).
  Other chains have their own scalings and regimes (not indicated in the figure).}\label{fig:PE_extension}
 \end{center}
\end{figure}

We then simulated polyelectrolyte (PE).
The settings are similar to the NC3 case except each monomer carries a negative unit charge $-e$.
$N_m$ monovalent cations were dissociated from the PE chain because of the electroneutrality.
A certain amount of monovalent salts were added into the solutions.
The ionic strength of solution, $I=\frac{1}{2} \sum_{i} z_{i}^2 c_{i}$, was kept at constant, 
where $z_{i}$ and $c_{i}$ are the valence and the concentration of the ion species $i$, respectively.
We assumed that an ion bead has mass $m$ and the WCA parameters $(\varepsilon_{\rm i},\sigma_{\rm i})=(1.2, 1.0)$.
The electrostatic interaction under periodic boundary condition was calculated using particle-particle particle-mesh Ewald  with the Bjerrum length $\bjerrum$ setting to $3.0$.
We studied the confinement of flexible PE chains at three ionic strengths: $I_{1}=0.00004$, $I_{2}=0.0004$ and $I_{3}=0.004$, which correspond to $5\,{\rm mM}$, $50\,{\rm mM}$, and $0.5\,{\rm M}$, respectively, in real units and cover from low to high ionic strength. 
The systems are denoted by PEfl-$I_1$, PEfl-$I_2$, and PEfl-$I_3$.
The results, shown in Fig.~\ref{fig:PE_extension}(a), were further compared with the semiflexible PE chain of $k_{\rm a}=3.5$ at $I=I_{1}$, denoted by PEsf-$I_1$, and the neutral chain NC3.

The behavior of $Z_{\rm ext}$ can be divided into de Gennes, backfolding and Odijk regimes.
Noticeably, the de Gennes regime is subdivided into two regimes, namely Flory-de Gennes regime and electro-de Gennes regime. 
When $D$ decreases from the bulk regime, the conformation of the PE chain is a connected Flory blobs.
It acquires the classical de Gennes behavior with $Z_{\rm ext} \sim D^{-0.67}$.
If $D$ becomes smaller than the Debye length $\lambda_\textmd{D}$, the electrostatic interaction dominates the blobs (called the electrostatic blobs). 
The subchains in the blobs are rodlike and the exponent $\nu$ is close to $1$.
A $D^0$-scaling behavior is anticipated. 
We found a real $D^{-0.12}$ variation for PEfl-$I_1$ from the simulations.
It gives $\nu \simeq 0.89$ according to the de Gennes relation $D^{1-1/\nu}$.
The boundary separating the Flory- and electro-de Gennes regimes locates at $D^*_{\rm ef} \sim 100$, $35$, $10$ for PEfl at $I_1$, $I_2$, $I_3$, respectively.
These boundary shifts follow essentially $I^{-1/2}$, a fact asserting $D^*_{\rm ef}\sim \lambda_\textmd{D}$.

In the region $3<D<7\sim 2\bjerrum$, the chain extension exhibits a $D^{-0.3}$-power law, no matter what the ionic strength is.
This confined space goes inside the region of ion condensation, generally described by a tube region of radius $\lambda_\textmd{B}$, surrounding the PE chain \cite{hsiao2006salt, grass2009polyelectrolytes}.
Inside it, the ionic screening does not function any more.
Consequently, the $Z_{\rm ext}$ curves collapse together for different $I$ values.
The calculation in Fig.~\ref{fig:PE_extension}(b) shows that nonzero $P_{\rm F}$ occurred as $D>3$. 
It enables to relate this region to the backfolding regime.
For $D<3$, the chains is in the Odijk regime.

Compared with the semiflexible chains, we observed a broader Odijk regime for PEsf-$I_1$.
Unlike NC1, there exhibits no transition regime.
It is because the electrostatic interaction dominates the chain stiffness in this case.
The high ionic strength behavior (PEfl-$I_3$) looks similar to the NC3 chain except the characteristics
of broader Odijk and backfolding regimes, owing to the electrostatic contribution to the persistence length. 

\begin{figure}[htbp]
\begin{center}
\includegraphics[width=0.30\textwidth,angle=270]{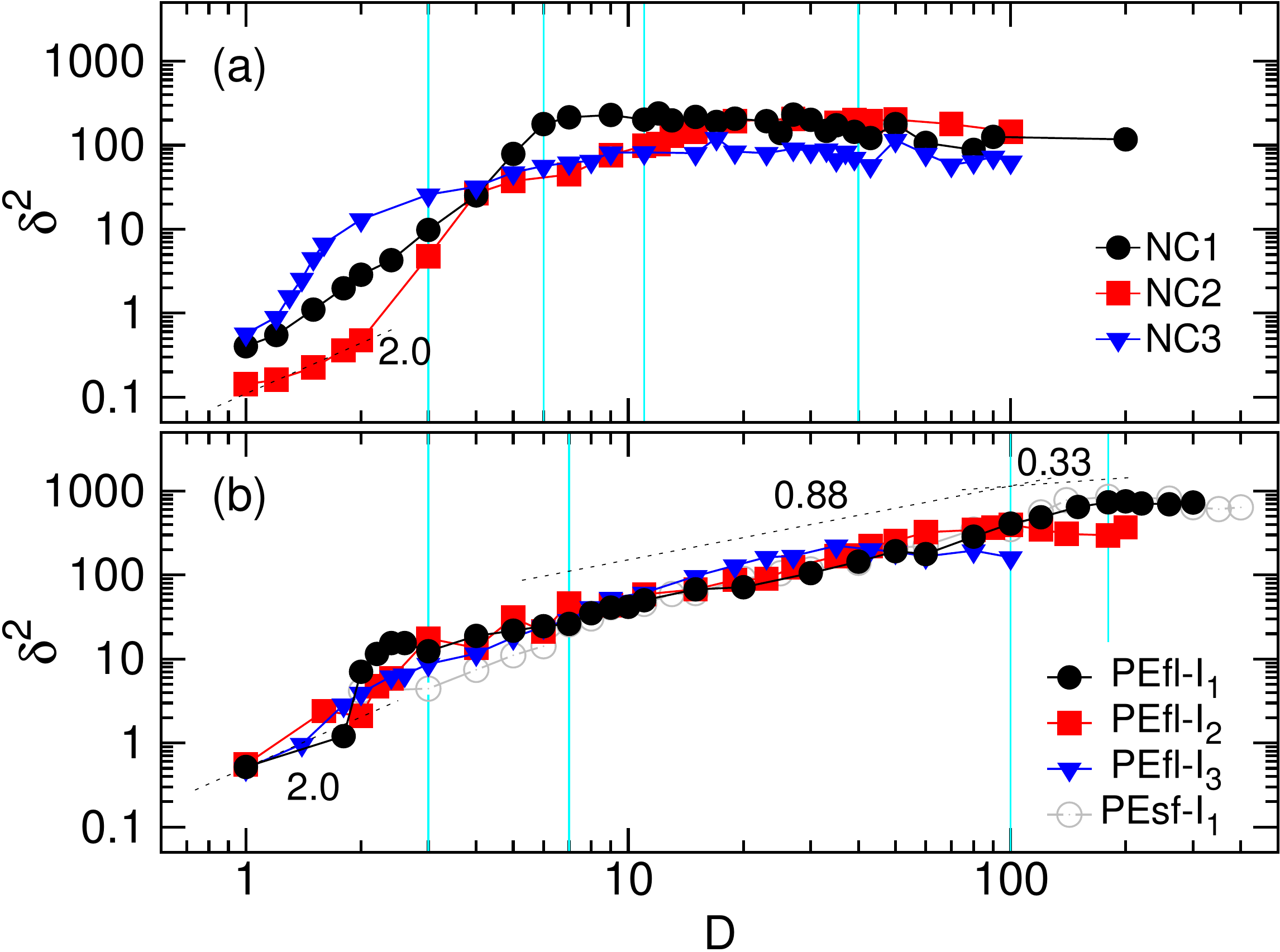}
\caption{Square of the fluctuation $\delta^{2}$ of $Z_{\rm ext}$ as a function of $D$ for (a) the NC chains and (b) the PE chains. 
The cyan vertical lines demarcate the regime boundaries for NC1 (in Panel (a)) and PEfl-$I_1$ (in Panel (b)).}\label{fig:SDV}
 \end{center}
\end{figure}

We studied furthermore the fluctuation of the chain extension, $\delta$. 
Theoretically,  $\delta^2$ is related to the thermal energy by $\frac{1}{2}k_{\rm eff}\delta^{2}\sim k_{\rm B}T$
where $k_{\rm eff}$ is the effective spring constant obtained from the 2nd derivatives of the chain free energy $F$ \cite{reisner2012dna, dai2016polymer}. 
In the channel, the polymer forms a 1D blob chain. 
Hence, $F$ is a sum of two terms: $F=\frac{N}{g_{\rm c}}F_{\rm bl}+F_{\rm bc}$, where $F_{\rm bl}$ is the free energy of a blob and $F_{\rm bc}$ is the one of the blob chain.
It leads $\delta^2=\frac{N}{g_{\rm c}}\Delta\xi_{\rm c\parallel}^2+\Delta R^2$.
For NC chains, the constituted blobs are soft and the fluctuations are mainly derived from the first term.
The predictions were obtained
\begin{eqnarray}
\delta^2 \sim 
\begin{cases}
 \frac{N\ell^2}{1 + A_1(\frac{\ell^{2}}{Nw^2})^{2\nu-1}} & \mbox{bulk,} \\[7pt]
 \frac{N\ell^2}{1 + A_2(\frac{\ell^{2}}{Dw})^{2-\frac{1}{\nu}}} & \mbox{de Gennes,} \\[7pt]
  N\ell^2 & \mbox{extended de Gennes,} \\
  \frac{N\ell^5 w}{v_{\rm ex}D} & \mbox{transition,}
\end{cases}\label{Eq-SDV}
\end{eqnarray}
where $A_1$ and $A_2$ are factors.
Fig.~\ref{fig:SDV}(a) presents the results of the simulations.
For NC1, $\delta^{2}$ is roughly maintained at a constant from the bulk to the extended de Gennes regime.
In the transition regime, it increases slightly with decreasing $D$, which becomes more obvious if the chain stiffness $k_{\rm a}$ is increased.
A similar fluctuation behavior has been reported in literatures~\cite{muralidhar2014backfolding}, in the region near $D\sim\ell$ when $\ell/w \gg 1$. 
These results are consistent with the predictions of Eq.~\ref{Eq-SDV}.
For NC2 and NC3, $\delta^{2}$ shows a smooth decrease, starting from a wider channel region because
both of the extended de Gennes and transition regimes are shrunken.

It has been argued that $\delta^{2}$ scales as $D^2$ in the Odijk regime using the deflection chain theory \cite{dai2016polymer}.
We did see this trend of behavior in the simulations.
If the chain enters the backfolding regime from the Odijk, the chain segments are allowed to fold in the backward direction.
The size fluctuation increases significantly, which causes the growth of $\delta^{2}$ faster than $D^2$.

Fig.~\ref{fig:SDV}(b) shows the calculations for PE chains.
$\delta^{2}$ decrease when the chain leaves the bulk regime.
Due to the strong electrostatic repulsion inside the blobs, the dominated contribution comes from the term of the blob chain $\Delta R^2$, which predicts $\delta^2 \sim N\ell^2 (\frac{Dw}{\ell^{2}})^{2-\frac{1}{\nu}}$ in the de Gennes regime.
The scaling passes $D^{0.33}$ and then arrives $D^{0.88}$, corresponding consistently to the Flory-de Gennes regime with $\nu\simeq 0.6$ and the electro-de Gennes with  $\nu\simeq 0.89$, respectively.
Below the two de Gennes regimes, the interplay between the electrostatic interaction and the excluded volume of segment renders $\delta^{2}$ a decreasing function, which follows by a chute, and finally shows $D^2$-scaling in the Odijk regime.

We have expanded the scaling theory and performed simulations to study polymers in different conditions of confinement.
Length scales such as the persistence length and the chain width for neutral polymers, together with the Debye length and the Bjerrum length for charged chains, exert their influences by working with the channel width, which results in a variety of the scaling behaviors.
When converting to real units, the newly predicted electro-de Gennes regime for flexible PE is expected to show in a channel of $D$ between $2.5$ and $25\,{\rm nm}$ at $I\simeq 5\,{\rm mM}$.
Increasing the chain stiffness and the ionic strength shrinks the regime from the lower and the upper boundaries, respectively.
Since the persistence length of dsDNA molecules is much larger than the upper boundary,  
the regime is not anticipated to exhibit in dsDNA confinement.
To observe the regime, flexible PE such as ssDNA or RNA should be used in experiments in a low ionic strength condition.

This material is based upon work supported by the Ministry of Science and Technology, Taiwan under the Contract Nos.~MOST 103-2112-M-007-014-MY3 and MOST 106-2112-M-007-027-MY3.  


\end{document}